\documentclass{iau}

\usepackage{amsmath}
\usepackage{graphicx}
\usepackage{multirow}

\UseRawInputEncoding      %REQUIRED FOR arXiv SUBMISSION
\newcommand{\Rsun}{~$R_\odot$}

\newcommand{\fov}{field of view}
\newcommand{\fovs}{fields of view}

\newcommand{\eg}{e.g., {\ }}

\begin{document}

\lefttitle{P. Lamy}
\righttitle{Observing Solar Coronal Mass Ejections from Space}

\jnlPage{1}{7}
\jnlDoiYr{2024}
\doival{10.1017/xxxxx}
\volno{388}
\pubYr{2024}
\journaltitle{Solar and Stellar Coronal Mass Ejections}

\aopheadtitle{Proceedings of the IAU Symposium}
\editors{N. Gopalswamy,  O. Malandraki, A. Vidotto \&  W. Manchester, eds.}

\title{Observing Solar Coronal Mass Ejections from Space}

\author{Philippe L. Lamy$^1$}
\affiliation{(1) Laboratoire Atmosph\`eres, Milieux et Observations Spatiales, CNRS \& UVSQ, Guyancourt, France}

\begin{abstract}
In this contribution to the panel discussion of the IAU Symposium N${^{\rm o}}$\,388 ``Solar and Stellar Coronal Mass Ejections'', I concentrate on white-light observations of solar coronal mass ejections (CMEs) from space and specifically address the following aspects: i) history of observations, ii) available catalogs of CMEs, iii) achievements of space observations of CMEs, iv) future of CME observation, and v) challenges and future directions. 
\end{abstract}


\maketitle

%===============================================================================
\section{Introduction}
%===============================================================================
Following the discovery of CMEs, there has been a rapidly growing interest in their observation and their analysis motivated by their role in solar, coronal, interplanetary, and planetary physics as well as in space weather with in particular, far reaching consequences for human activities on Earth.
Successive space solar missions have been implemented with among their objectives, the requirement to maintain a continuous monitoring of CME activity.
As a consequence, the number of investigations of CMEs over the past four decades has exploded: several thousand refereed publications and at least 80 review articles, and we now see a rapidly growing field of star CMEs.
It is obviously beyond the scope of this concise presentation of the space observations of CMEs to quote this extensive literature, but the interested reader is directed to recent reviews (\eg Lamy et al.\ 2019).
This presentation is limited to coronagraphic observations of CMEs and organized in five sections which briefly address the following aspects: i) history of observations, ii) available catalogs of CMEs, iii) achievements of space observations of CMEs, iv) future of CME observation, and v) challenges and future directions. 

%===============================================================================
\section{Brief History of CME Observations}
%===============================================================================
After the first detection of a CME by the coronagraph onboard the seventh \textit{Orbiting Solar Observatory} (OSO-7) in late 1971, the history of space observations of CMEs may be conveniently divided in three periods.
The early years from 1971 to 1995 saw the first generation of space-borne coronagraphs: Skylab-ATM, P78-1 Solwind, and the Coronagraph/Polarimeter (C/P) on the \textit{Solar Maximum Mission} (SMM).
In January 1996, the Large Angle Spectrometric Coronagraph (LASCO) onboard the \textit{Solar and Heliosphere Observatory} (SoHO) opened a golden age for CMEs, later complemented by the COR2 A and B coronagraphs onboard the STEREO twin spacecraft which offered two viewpoints until 2014 (STEREO-B was then lost) and an enlarged \fov\ thanks to the inclusion of Heliospheric Imagers (HIs).
The LASCO-C2 and -C3 remain the workhorses for CMEs investigations and for space weather forecasts with a record of quasi uninterrupted observations over 2.5 solar cycles (and hopefully three) in the range 2.5\,--,30\,\Rsun.
The third stage is characterized by the emergence of new players.
The Wide Field Imager for Solar Probe (WISPR) onboard the \textit{Parker Solar Probe} (PSP) offers for the first time views from the interior of a few CMEs.
The METIS ultraviolet+visible coronagraph onboard the \textit{Solar Orbiter} has however its operation limited to restricted observational windows, thus precluding continuous monitoring. 
Consequently, only a handful of CMEs have been detected and analyzed so far.
The SCI coronagraph onboard the \textit{Advanced Space-based Solar Observatory} (ASO-S) is likewise METIS an ultraviolet+visible instrument, but with a much narrower \fov.
SCI has observed quite a number of CMEs, but these observations are not yet documented.

%===============================================================================
\section{Catalogs of CMEs}
%===============================================================================
The very large number of CMES seen by LASCO soon imposed the necessity of their systematic detection and of a compilation of their properties.
Following a first attempt performed by St. Cyr et al. (2000) who reported the properties of 841 CMEs from January 1996 to June 1998, the Coordinated Data Analysis Workshop catalog (CDAW) was initiated based on the visual detection of CMEs by Yashiro et al. (2004), see also Gopalswamy et al. (this volume) for a second version.
Biases inherent to visual detection were pointed out in the literature (\eg Webb \& Howard\ 2012), leading to the development of a new generation of catalogs based on automatic detection.
Four catalogs are presently available implementing different detection techniques: CACTus (Robbrecht et al.\ 2004), SEEDS (Olmedo et al.\ 2005), ARTEMIS (Boursier et al.\ 2009; Floyd et al.\ 2013), and CORIMP (Byrne et al.\ 2009) however limited to the [2000\,--\,2015] time interval.
They all report the following basic properties: date of CMEs first appearance in C2, central position angle, angular width, speed (one or several).
CDAW, SEEDS, and CORIMP include acceleration.
These three catalogs, together with ARTEMIS, include mass and kinetic energy.
Whereas automated methodologies are more objective, the listed properties are often inconsistent with each other and furthermore with the CDAW manual catalog.
As an example, SEEDS reports about twice as many CMEs as CDAW, the other catalogs (except CORIMP) falling in-between. 
This is not a surprise in view of the intrinsic difficulty of identifying and characterizing CMEs which display a large diversity of apparent shapes. 
This situation clearly requires a complementary use of all catalogs when studying particular CMEs in order to ascertain their results. 
However, in their extensive comparison of the different catalogs, Lamy et al. (2019) pointed out that there are also many points of convergence, particularly between SEEDS (once corrected for the image cadence) and ARTEMIS.

The CACTus and SEEDS catalogs have been adapted to incorporate the STEREO/COR2 data starting in 2017.
In addition, the new MVC catalog was built from seven years of multi-viewpoint observations listing 3358 events simultaneously observed by the twin COR2 coronagraphs (Vourlidas et al.\ 2017).
Finally, a catalog based on WISPR observations is in preparation.

%===============================================================================
\section{Achievements of Space Observations of CMEs}
%===============================================================================
Over the past decades, space-based observations of CMEs have led to numerous significant scientific achievements, revolutionizing our understanding of these complex and dynamic events. 
A synthetic list of the main scientific achievements is presented below.
\begin{itemize}
\item
CMEs and solar cycle variability: Space-based observations have revealed that CME rates and properties vary significantly over the solar cycle, with implications for our understanding of solar cycle variability and its impact on the heliosphere.
The CME occurrence rate closely tracks solar activity (Fig. 1 and 2) and particularly the solar radio flux at 10.7\,cm, probably because it is a coronal index (although it is partly chromospheric) rather than a photospheric index.
\item 
CMEs energy and helicity: Space-based observations have fostered our understanding of how magnetic energy is built up, stored, and released in magnetic flux systems and of the role of CMEs in removing large amounts of magnetic flux and helicity from the Sun, thus making room for the new solar cycle.
\item 
CME initiation and acceleration mechanisms: Space-based observations have provided insights into the initiation and acceleration mechanisms of CMEs. 
For example, they have revealed the importance of magnetic reconnection and flux rope formation in CME initiation.
\item 
CME associations: Space-based observations have revealed that only a small fraction (a few percent) of the overall population of CMEs is associated with C+M+X class flares and erupting prominences.
These associated CMEs have significantly larger mass, speed, and kinetic energy.
The bulk of the CME population is associated with streamers.
\item
Improved CME detection and tracking: Space-based observatories have enabled the detection and tracking of CMEs with unprecedented accuracy. 
This has led to a better understanding of CME dynamics, kinematics, and morphology.
\item 
CME-CME interactions and merging: Space-based observations have shown that CMEs can interact and merge with each other (``cannibalism''), leading to complex and unpredictable behavior. 
This has significant implications for space weather forecasting and geomagnetic storm prediction.
\item
CMEs and solar energetic particle (SEP) events: Space-based observations have established a strong connection between CMEs and SEP events. 
This has led to a better understanding of the acceleration and transport of high-energy particles in the solar corona and interplanetary space.
\item 
CME-driven shocks and radio bursts: Space-based observations have revealed the importance of CME-driven shocks in generating radio bursts and type II radio emissions. 
This has significant implications for our understanding of CME dynamics and their impact on the solar wind and magnetosphere.
\item 
CMEs and geomagnetic storms: Space-based observations have demonstrated the critical role of CMEs in driving geomagnetic storms, which can have significant impacts on Earth's magnetic field, radiation belts, and upper atmosphere.
\item 
CMEs and space weather forecasting: Space-based observations have enabled the development of CME forecasting models, which are critical for predicting space weather events and mitigating their impacts on Earth's magnetic field, radiation belts, and technological systems.
\end{itemize}

Altogether, these achievements have significantly advanced our understanding of CMEs and their role in shaping the solar corona, heliosphere, and space weather.

\begin{figure}[htpb!]
\begin{center}
%\hspace*{0.5cm}\includegraphics[angle=0,scale=0.35]{shen_et12_event.jpg}
\includegraphics[width=\textwidth]{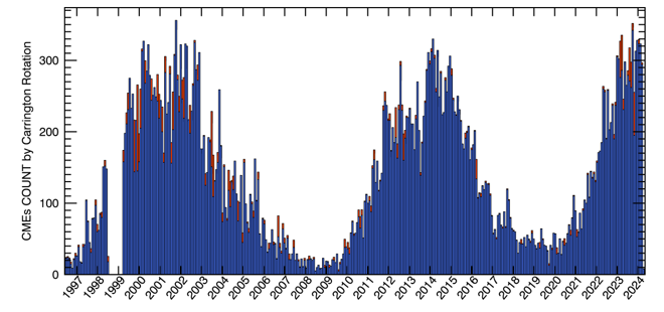}
\caption{Monthly detection rate of CMEs reported in the ARTEMIS catalog (in blue) further corrected for the duty-cycle of LASCO (in red).
Note the data gap in 1998 when SoHO went out of control.}
\end{center}
\end{figure}

\begin{figure}[htpb!]
\begin{center}
\includegraphics[width=\textwidth]{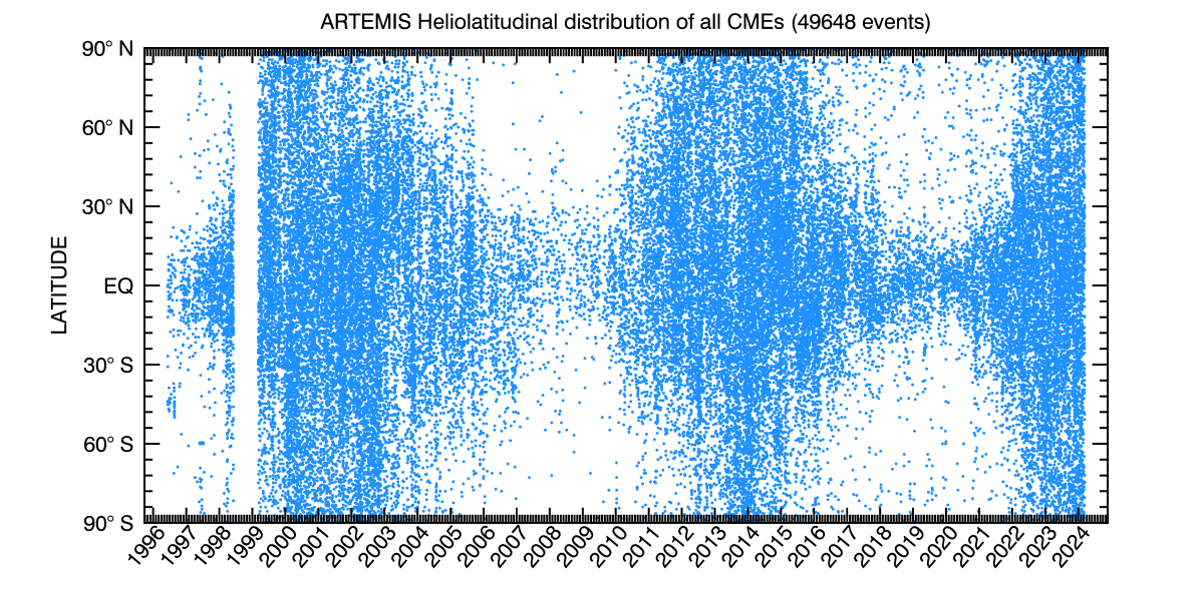}
\caption{Heliolatitudinal apparent distribution of CMEs reported in the ARTEMIS catalog.
Note the data gap in 1998 when SoHO went out of control.}
\end{center}
\end{figure}

%===============================================================================
\section{Future Observations of CMEs}
%===============================================================================
There is a substantial number of solar space missions that have recently been launched or that will be launched in the near future.
They are listed below in chronological order.

\begin{itemize}
\item 
India Aditya-L1/VELC: launched in September 2023 at L1 Lagrangian point, this emission line coronagraph with a \fov\ of 1.05\,--\,3 \Rsun\ is crippled by significant straylight problems probably precluding CME observations.

\item 
NOAO GOES-U/CCOR-1: scheduled to launch in June 2024 on a geosynchronous orbit, this compact basic coronagraph without polarization capability has a \fov\ of 3.7\,--\,17 \Rsun.
As a consequence of is orbit, the Earth will approach and or even transit its \fov\ every day thus limiting its operation.  

\item 
ESA Proba-3/ASPIICS : scheduled to launch in November 2024 on a highly elliptical Earth orbit, this will be the first ``giant'' externally occulted coronagraph comprising two distinct satellites (occulter and instrument) separated by 150~m in formation flight .
It has a \fov\ of 1.1\,--\,3 \Rsun\ and polarization capability and will operate typically six hours per day.

\item 
NOAO SVFO/CCOR-2: scheduled to launch in mid 2025 at L1, this compact basic coronagraph is similar to CCOR-1 except for a \fov\ of 3.0\,--\,22 \Rsun. 

\item 
NASA Punch: scheduled to launch in April 2025 on a Sun-synchronous, low Earth orbit, it comprises four satellites, one with a coronagraph having a \fov\ of 6.0\,--\,32 \Rsun\ and three with heliospheric imagers having a \fov\ of 18\,--\,180 \Rsun.
These instruments have polarization capability.

\item 
ESA Vigil/CCOR-3  (ex Lagrange): scheduled to launch in 2031 at L5, this satellite will embark a compact basic coronagraph similar to CCOR-2, that is a \fov\ of 3.0\,--\,22 \Rsun\ and no polarization capability.

\end{itemize}

%===============================================================================
\section{Challenges, Future Directions, and Conclusion}
%===============================================================================
Whereas there is a substantial number of solar space missions that will target CMEs as seen in the above section, many of them implement very simple  coronagraphs at a single viewpoint to satisfy the basic need of space weather forecasting.
To really make progress in the understanding of CMEs and addressing outstanding questions such as their initiation, their internal structure and their energetics, a comprehensive mission with global science objectives, multiple viewpoints, and continuous imaging over large \fovs\ is required.
Several missions that combine observations from several of the Lagrangian points L1, L4, and L5 have been proposed, but were unsuccessful so far.

The newly proposed MOST mission (``Multiview Observatory for Solar Terrestrial Science'', Gopalswamy et al.\ 2024) is intended to move forward and to make substantial progress with respect to past missions.
Briefly, it consists of four spacecraft at L4, L5, ahead of L4, and behind L5. 
The first two will have identical remote-sensing and in-situ instrument suites including coronagraphs and heliospheric imagers. 
Altogether, the four spacecraft will perform novel Faraday rotation measurements to measure the magnetic content of CMEs.
Assuming an additional solar observatory at L1, highly accurate three-dimensional information will be obtained.

This is of utmost importance as detecting, tracking, and characterizing CMEs remain extremely challenging as illustrated by the large uncertainties in the measured parameters, especially in the case of a single viewpoint, as pointed out in Section~2. 
Deep learning looks extremely promising to make progress and several efforts are on-going.
For instance, it has been implemented by Shan et al. (2024) to produce an automated ``2D CME'' catalog from one year of STEREO/COR1-A observations.
Further applying the polarization technique, they upgrade it to a ``3D CME'' catalog that includes 3D propagation direction, speed, and 3D shape reconstruction.
This clearly demonstrates the importance of polarization measurements for the 3D characterization of CMEs.

Outside the scope of this presentation, but worth mentioning, MHD modeling is fundamental to analyze and interpret the observations of CMEs.
Several models are presently available, but improvements are needed to foster our understanding of their complexity.

\end{document}